# Nonlinear wavelength conversion in photonic crystal fibers with three zero dispersion points

**S. P. Stark, F. Biancalana, A. Podlipensky and P. St.J. Russell**

Max Planck Institute for the Science of Light

Guenther-Scharowsky Str. 1/Bau 24, 91058 Erlangen, Germany

http://mpl.mpg.de

In this theoretical study, we show that a simple endlessly single-mode photonic crystal fiber can be designed to yield, not just two, but three zero-dispersion wavelengths. The presence of a third dispersion zero creates a rich phase-matching topology, enabling enhanced control over the spectral locations of the four-wave-mixing and resonant-radiation bands emitted by solitons and short pulses. The greatly enhanced flexibility in the positioning of these bands has applications in wavelength conversion, supercontinuum generation and pair-photon sources for quantum optics.

## 1. Introduction

Photonic crystal fibers (PCFs) [1] have opened new opportunities for effectively manipulating the properties of optical radiation. Solid-core PCFs, in which light is guided by conventional total internal reflection (TIR), have proved their usefulness in many nonlinear applications [see e.g. 1 - 3]. The large practical success of the PCF concept stems largely from the fact that the dispersion properties of the fibers can be engineered to a very high degree [4].

For example, it is well-known that the correct dispersion characteristics are pivotal for obtaining phase-matching (PM) in the majority of nonlinear effects in fiber optics. Four-wave mixing

(FWM) of continuous wave (CW) laser light [5], and so-called resonant "Čerenkov" radiation [6, 7] emitted by ultrashort solitons, depend crucially on the frequency dependence of the propagation constant β. Both effects lead to the exponential ('parametric') growth from noise of new phase-matched frequencies, gain being provided by the pump wave. Degenerate FWM is a nonlinear process where two pump photons are annihilated and two new photons are created, symmetrically detuned from the pump by a frequency-shift Ω [8]. Resonant radiation (RR) on the other hand is typically an asymmetric process, requiring the existence of a short pulse whose spectral components lie on a straight line in ω-k space, for example a fiber-optical soliton [9].

Fibers with one zero-dispersion wavelength (ZDW) in their group velocity dispersion (GVD) profile, for example large-core fibers, have been extensively studied. FWM is always phase-matched when the GVD is anomalous [8]. Additionally, non-vanishing higher-order dispersion coefficients allow for PM in narrow bands within the normal GVD region [5]. Resonant radiation has been identified as an important ingredient in octave-spanning spectral broadening [10]. With the advent of PCF, the fabrication of solid core fibers with two ZDWs has become a reality. These fibers, like nonlinear PCFs and tapered fibers, have turned out to be a rich playground for nonlinear studies. The FWM topology in these fibers shows phase-matched excitation of up to four sidebands [11], which has proved useful for applications in quantum optics [12]. In addition, the second ZDW has a huge impact on the propagation of ultrashort pulses. It can lead to the emergence of two bands of phase-matched resonant radiation which shape the SC spectrum uniquely and allow the redistribution of the pump energy with high efficiency [13].

## 2. Dispersion characteristics

Some years ago it was shown that the structure of small-hole endlessly single-mode PCF can be tailored to provide an ultraflat GVD profile [4]. In this Letter we show that this type of PCF can also exhibit *three* ZDWs, with the result that the PM curves not only exhibit additional phase-matched frequencies, but have a complicated algebraic topology.

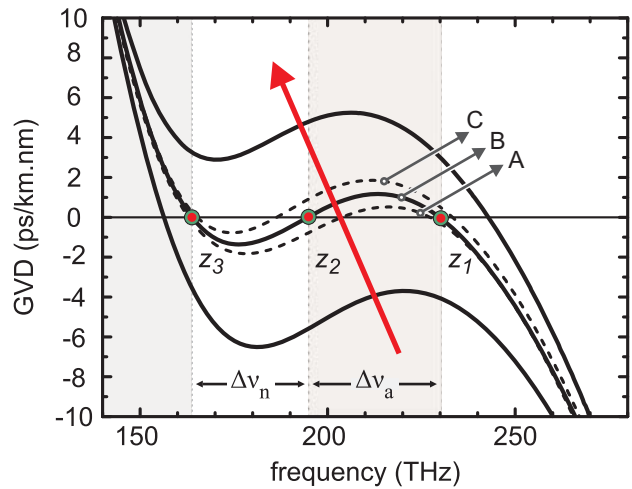

**Fig. 1.** (Color online) Dispersion profiles for different inter-hole spacings at hole-to-pitch ratio $d/\Lambda = 0.12$. The red arrow marks the general trend for increasing pitch (from 2.5 μm to 2.7 μm). A-C labels the curves used for the PM calculations, with 2.6 μm (A), 2.612 μm (B) and 2.62 μm (C). The shading indicates the anomalous dispersion regions and curve B is used to define the widths $\Delta v_a$ and $\Delta v_n$.

In Fig. 1 we show numerical calculations of GVD profiles, obtained using the fixed-frequency plane wave (FFPW) method [1]. The fiber geometry considered in this Letter follows closely the structure of the fiber reported in [4], where it was experimentally shown that the dispersion curve is very flat and small in magnitude over a broad wavelength region. By carefully choosing the inter-hole spacing at a hole diameter of 600 nm we have found that the GVD curves have *three* ZDWs (see Fig. 1). In the upper right inset of Fig. 1 are the GVD curves used in the following PM study. We will show that the PM wavelengths depend on the relative degree of asymmetry between the widths $\Delta v_a$ and $\Delta v_n$ of the anomalous and normal dispersion regions, parameterized by $\alpha = (\Delta\nu_n - \Delta\nu_a)/\Delta\nu_n$. The value of α is very sensitive to the pitch, and can be both positive and negative (see Fig. 1).

## 3. Four-wave mixing

In degenerate FWM, energy conservation determines the frequencies of the signal (anti-Stokes, frequency $\nu_{AS}$) and idler (Stokes, frequency $\nu_S$) waves generated from the pump light (frequency $\nu_0$). The process is most intense when wave momentum is conserved. Both requirements translate into [8]:

$$\begin{aligned}2\nu_0 - (\nu_S + \nu_{AS}) &= 0 \\ \beta(\nu_S) + \beta(\nu_{AS}) - 2\beta(\nu_0) + 2\gamma P_0 &= 0\end{aligned} \quad (1)$$

where γ is the nonlinear coefficient of the fiber and $P_0$ is the pump power. We now present PM curves for FWM processes in fibers with 3 ZDWs and show that their topology has several unique features.

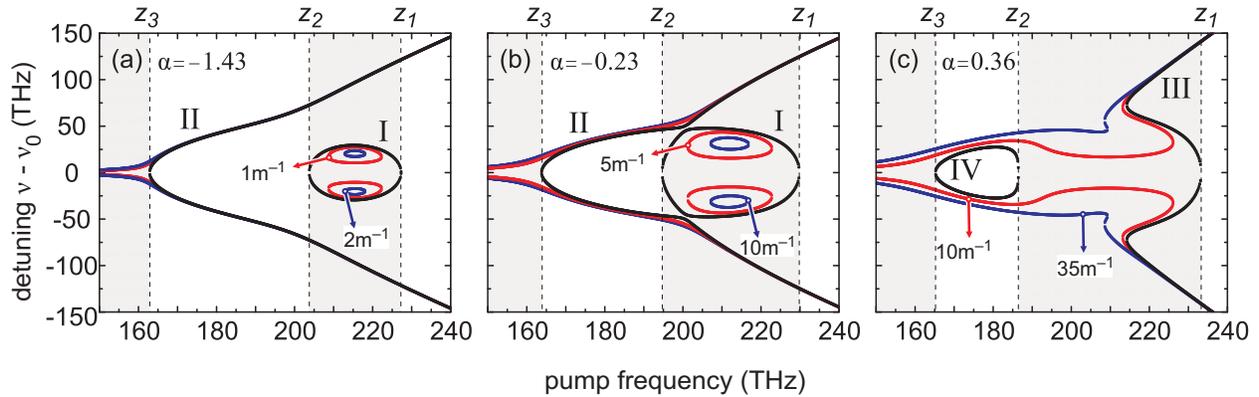

**Fig. 2.** (Color online) Topology of PM curves for fibers with three ZDWs. (a) corresponds to fiber A, (b) to fiber B and (c) to fiber C shown in Fig. 1. The black curves are for γP = 0. The other curves are labelled by their corresponding value γP. We used $\gamma = 10$ W$^{-1}$km$^{-1}$, obtained with the FFPW method. The shaded regions have anomalous dispersion.

Fig. 2 shows the PM curves for different values of α and input pump power. We find that the PM curves can provide parametric gain at up to *six* frequencies, for certain pump frequencies within $\Delta v_a$. In addition, the FWM PM topology depends on the parameter α.

In Fig. 2a we show the PM for α = −1.43 (i.e., $\Delta v_n > \Delta v_a$) which is associated with the GVD curve A in Fig. 1. The topology features two PM curves, labelled **I** and **II**. The PM curve **I** is confined between $z_1$ and $z_2$ and can provide gain to up to four PM sidebands, depending on the frequency and power. In contrast to this behaviour, the newly arising third (long-wavelength) ZDW $z_3$ causes the formation of an additional PM branch **(II)**, which is phase-matched over a broad frequency range. It extends beyond the high frequency anomalous GVD region, bringing the total number of PM sidebands between $z_1$ and $z_2$ up to a total of six. When the pump power $P_0$ exceeds the critical value $3\beta_2/(2\beta_4\gamma)$, only branch **II** is phase-matched and branch **I** vanishes.

When the spacing between $z_1$ and $z_2$ falls, the two PM curves approach each other until they finally touch, as can be seen in Fig. 2b (α = −0.23). A further increase of α leads then to *avoided crossings* of the PM curves, creating a new topology. We demonstrate this in Fig. 2c, where the fiber dispersion profile is such that $\Delta v_n < \Delta v_a$ (α = +0.36), and the avoided crossings result in the appearance of two new PM curves, labelled **III** and **IV.** The frequency regions for which the pump photons can generate six sidebands are now strongly limited, appearing near $z_3$. It is very interesting to see that in the limit of vanishingly small power (black curve) the PM curve **IV** supplies gain for all frequencies in $\Delta v_n$.

## 4. Numerical modelling

In the following we support our findings by studying the propagation of near-CW pulses. The governing equation is the generalized nonlinear Schrödinger equation:

$$\frac{\partial A(z,\tau)}{\partial z} = \hat{D}A(z,\tau) - \frac{\alpha}{2}A(z,\tau) + i\left(\gamma(\omega_0) + i\gamma_1\frac{\partial}{\partial \tau}\right) \times \left(A(z,\tau)\int_{-\infty}^{\infty}R(t')|A(z,\tau)|^2 dt'\right) \quad (2)$$

where $A(z, \tau)$ is the complex time-domain envelope of the electric field, $\tau = t - z/v_g$ is the time in a reference frame moving with the group velocity of the pulse, $t$ the physical time and $v_g$ the group velocity. The operator $\hat{D}$ takes care of dispersion – for more details refer to [13]. In our simulations we study the dynamics of 50 ps pulses, long enough to limit dispersive effects in the pump pulses and to prevent temporal walk-off from the generated sidebands. The fiber we used has zero dispersion points at 165.2 THz, 186.5 THz and 233.3 THz (see Fig. 1, fiber C). In Fig. 3 we show the spectral evolution along the fiber length for light launched at different frequencies. In Fig. 3a the input frequency is $v_0 = 220.4$ THz and, from Fig. 2c, six PM sidebands are expected. This is confirmed in the simulations, six sidebands appearing after 2 m of propagation. When the pump frequency is increased to $v_0 = 228.8$ THz, only two PM sidebands are expected, a prediction that is confirmed in Fig. 3b, the frequencies of the two sidebands exactly corresponding with the PM calculations.

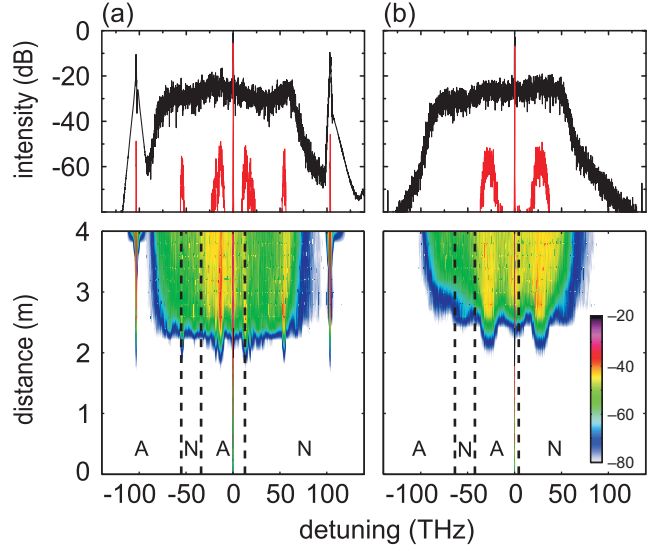

**Fig. 3.** Simulated propagation of 50 ps pulses in the fiber C (see Fig 1). The lower plots show the spectral evolution along the fiber length. The input spectrum (zero detuning) is too narrow to be clearly visible. The upper plots show the spectrum after 2 m (red) and 4 m (black). In (a) $v_0 = 220.4$ THz, resulting in 6 PM sidebands as expected from Fig. 2c. (b) Increase of the input frequency to $v_0 = 228.8$ THz, showing only two sidebands. The color scale is in decibels. The dashed lines indicate the ZDWs, separating the anomalous (A) and the normal (N) dispersion region.

## 5. Resonant radiation

We investigate the ability of 3 ZDW PCFs to modify the propagation of ultrashort pulses and to generate light at new frequencies. In general, solitons with peak amplitude $\sqrt{P}$ adapt to external perturbations, such as higher-order dispersion, by acquiring an oscillating tail, the resonant radiation (RR). This radiation builds up constructively when the resonance PM condition $\beta(v) - \beta(v_0) - \beta_1(v_0)(v - v_0) = \gamma P / 2$ is satisfied. This PM condition differs from the one for parametric FWM in that the sidebands are no longer restricted to symmetric positions on both sides of the pump frequency. Although bright solitons only exist when the GVD is anomalous, we find it very helpful to present the RR topology over the full range between 150 and 240 THz, including both normal and anomalous regions.

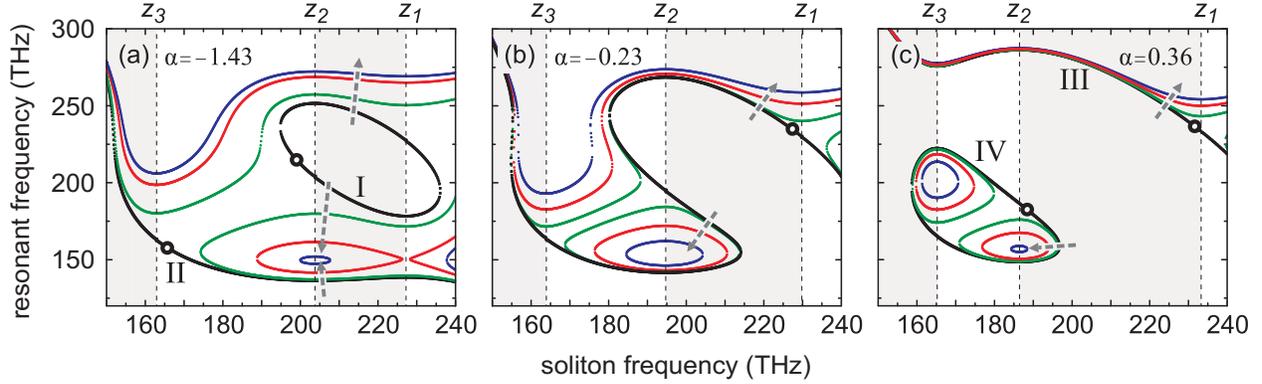

**Fig. 4.** (Color online) Resonant radiation topology for fibers with three ZDWs. The dashed arrows indicate how the curves change for increasing power. Curves for $\gamma P_0 = 0$ are marked with a black circle. Calculations based on: (a) dispersion profile A in Fig. 1 for $\gamma P_0 = 0$, 10, 50 and 70 m$^{-1}$; (b) dispersion profile B in Fig. 1 for $\gamma P_0 = 0$, 1, 10 and 25 m$^{-1}$; (c) dispersion profile C in Fig. 1 for $\gamma P_0 = 0$, 1, 5 and 10 m$^{-1}$. The shaded regions have anomalous dispersion.

Fig. 4 shows the RR topology for increasing values of α (from (a) to (c)), for the same PCFs as in Fig. 2. In general, regions can be identified where solitons are able to phase-match to three resonant frequencies. A change in the parameter α or the power dramatically modifies the RR topology.

### 5.1.1 Asymmetry α = −1.43

For this value of asymmetry the topology consists of two curves, labelled **I** and **II**, which are clearly separated in the limit $P \to 0$ (the black curves in Fig. 4a). In this limit the PM curve **I** is localized between $z_1$ and $z_2$, allowing PM to two resonances. The existence of the third zero dispersion point adds the additional PM branch **II** to the topology, which is not limited to certain frequencies but extends into the anomalous GVD region between $z_1$ and $z_2$. This means that solitons which are located in $\Delta v_a$ can shed energy into three dispersive waves. In contrast, solitons in the low-frequency anomalous dispersion region can phase-match to only one resonance.

When the soliton peak power is increased, the topology changes strongly (the green, red and blue curves in Fig. 4a). At first, PM curve **I** expands since the rising power causes a larger detuning of the RR from the pump frequency. Thus the low frequency sections of curves **I** and **II** approach each other. For sufficiently high power, the two curves **I** and **II** create an avoided crossing. Because of this, further increase in the power reduces the frequency range where solitons can phase-match to three RR frequencies (see the blue curve).

### 5.1.2 Asymmetry $\alpha = -0.23$ and $0.36$

In Figs. 4b&c we show how increasing α changes the RR topology. The excitation of three RR bands is now feasible for solitons in both anomalous GVD regions, in contrast to the case in Fig. 4a. For example, for $\alpha = -0.23$, a narrow frequency range can be identified in the high frequency anomalous GVD region where solitons can phase-match to three RR bands, all blue-detuned from the pump frequency. Around $z_1$, however, solitons can phase-match to only one RR band. For even higher pump power in the high frequency anomalous GVD region, the topology resembles the one shown in Fig. 4a and solitons can radiate into only one RR band.

Moving $z_2$ closer to $z_3$ leads to further modifications in the PM diagram. For $\alpha = 0.36$ (Fig. 4c) two new PM curves appear, labelled **III** and **IV**. In this case broad regions exist in both anomalous GVD regions where solitons can shed energy into three RR bands. Increasing the soliton peak power has a strong effect on curve **IV**; it splits, frequency separation between soliton and RR steadily decreasing.

### 6. Ultrashort pulse dynamics in 3 ZDW PCF

RR can cause pronounced reshaping of the spectrum during soliton propagation. The introduction of a third ZDW opens up new opportunities for modulating ultrashort pulses in both temporal and spectral domains. Here we study the behaviour of sech$^2$-pulses with carrier frequency set to 200 THz, within the high frequency anomalous GVD region of fiber C. We have chosen this particular set of input conditions to demonstrate the complex ultrafast dynamics resulting from the RR topology in Fig. 4c.

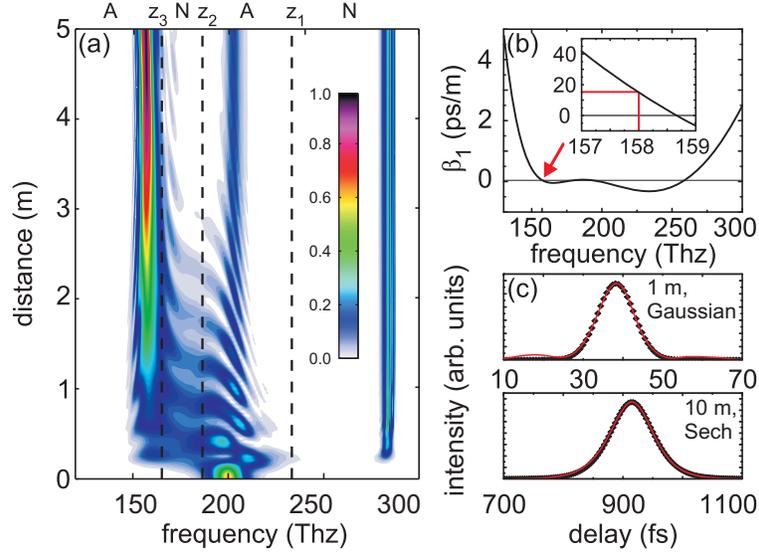

**Fig. 5.** Simulated propagation of 25 fs pump pulses in the fiber C (see Fig 1). (a) Spectral evolution along the fiber length; the dashed lines mark the three ZDWs and the pump frequency was 200 THz. (b) Inverse group velocity of the fiber; the inset zooms in on the region around 158 THz (the vertical scale is fs/m). (c) The IR pulse after 1 m (upper) and 10 m (lower) of propagation. The dashed lines indicate the ZDWs ($z_i$), separating the anomalous (A) and the normal (N) dispersion region.

For this carrier frequency only one RR band is phase-matched. The numerically modelled spectral evolution with propagation distance (Fig. 5) shows a dramatic redistribution of the pulse energy to new frequencies. Initially, after the launch of the pulses, a radiation band appears at 284 THz as expected by the PM calculations. However, due to the Raman self-frequency red-shift and energy dissipation the soliton moves into a parameter space where it is able to emit radiation into the IR. In Fig. 4c we see, that the PM curve **IV** can suddenly be phase-matched when the soliton frequency lowers. This effect is visible in our pulse propagation simulations, resulting in strong energy transfer into the IR.

Within the central region of anomalous dispersion, the light oscillates up and down both spectrally and spatially, creating a pattern of slanting fringes. The origin of this can be understood by means of a simplified model in which two pulses with constant carrier frequencies (corresponding to phase-matched soliton and RR bands at $\omega_1/2\pi$ = 197 THz and $\omega_2/2\pi$ = 158 THz in Fig. 5), walk off from each other temporally at a constant rate. This accumulating temporal delay $\Delta T(z)$ leads to a z-dependent change in relative phase between the pulses, given by $\Delta\phi = (\kappa_2 - \kappa_1)z + (\omega_2 + \omega_1)\Delta T(z)/2$ where $\kappa_1$ and $\kappa_2$ are the average propagation constants of the two pulses. During propagation, new spectral lines progressively accumulate at a typical length scale $L_B$, forming a modulated spectral comb in the anomalous band between $z_1$ and $z_2$ in Fig. 5 and leading to a recurrence at the pump frequency when $\Delta\phi$ increments by integer multiples of $2\pi$. Given that phase-matching requires $\kappa_1 = \kappa_2$, and that $\Delta T$ is related to the inverse group velocity difference between the pulses, i.e., $\Delta T = z \cdot (\beta_1(\omega_2) - \beta_1(\omega_1)) = z \cdot (v_{g2}^{-1} - v_{g1}^{-1})$, the spatial beat length of the pattern is:

$$L_B = \frac{4\pi}{(\omega_2 + \omega_1)(\beta_1(\omega_2) - \beta_1(\omega_1))}. \qquad (3)$$

For the parameters in Fig. 5 we have $\beta_1(\omega_2) - \beta_1(\omega_1) \simeq 15$ fs/m and Eq. (3) predicts $L_B$ = 37.5 cm, which is close to the ~35 cm period seen in the numerical simulations (Fig. 5a); note that this value is very sensitive to the precise values of frequency.

The IR pulse carries more than 88% of the input energy and shows no pronounced change in width or duration over several m of propagation. After its appearance, the RR pulse in the IR is Gaussian in shape with duration 10.75 fs and energy 13.2 pJ (Fig. 5c). During propagation, the pulse disperses, finally forming a fundamental soliton. This is shown in Fig. 5d, where the pulse fits well to a sech$^2$ envelope with duration 48 fs and energy 12.8 pJ. Since the central pulse frequency has not shifted, the pulse corresponds to a soliton of order $N = 1.5$. For longer propagation the pulse undergoes the usual soliton self-frequency shift (not shown). This effect has been termed spectral soliton tunnelling [14, 15], and the mechanism observed in our simulations agrees well with the findings in [15].

## 7. Conclusions

In this Letter we have explained how the presence of three ZDWs in special types of endlessly single-mode PCF produces unconventional phase-matching topologies for FWM and resonant radiation. This results in highly unusual supercontinuum spectra, soliton dynamics and resonant radiation properties. Multiple FWM peaks appear, the location of which can be effectively manipulated by slightly changing the relative position of the three ZDWs. These new peaks are likely to be of great interest in quantum optics, since the correct design of the fiber allows to optimally place the sidebands in frequency regions (far from the pump) where the Raman scattering is less important, thus reducing the impact of Raman decorrelations and leading to an increase of the rate of entangled photons in PCFs. The fabrication of PCFs with three ZDWs will require control of the PCF microstructure to better than 1%, which is within the tolerances of the drawing process.

## 8. REFERENCES

1. P. St.J. Russell, J. Lightwave Technol. **24**, 4729 (2006).
2. J. K. Ranka et al., Opt. Lett. **25**, 25 (2000).
3. J.M. Dudley and J.R. Taylor, Nature Photonics **3**, 85 (2009).
4. W.H. Reeves *et al.*, Nature **424**, 511 (2003).
5. F. Biancalana et al., Phys. Rev. E **68**, 046603 (2003).
6. F. Biancalana et al., Phys. Rev. E **70**, 016615 (2004).

7. N. Akhmediev and M. Karlsson, Phys. Rev. A **51**, 2602 (1995).
8. G.P. Agrawal, *Nonlinear Fiber Optics*, 4th Edition ed. (Academic, 2007).
9. A. Hasegawa and M. Matsumoto, *Optical Solitons in Fibers* (Springer, Heidelberg, 2003).
10. J. Herrmann et al., Phys. Rev. Lett. **88**, 173901 (2002).
11. A.Y.H. Chen et al., Opt. Lett. **30**, 762 (2005).
12. C. Söller et al., Phys. Rev. A **81**, 031801 (2010).
13. S. Stark et al. J. Opt. Soc. Am B **27**, 592 (2010).
14. V. N. Serkin et al., Electron. Lett. **29**, 12 (1993).
15. E.N. Tsoy and C.M. de Sterke, Phys. Rev. A **76**, 043804 (2007).